\documentclass[aps,nofootinbib,prd,preprintnumbers,showpacs]{revtex4}
\usepackage{wrapfig}
\usepackage{amsfonts}
\usepackage{amsmath}
\usepackage{hyperref}
\usepackage{amssymb}
\usepackage[english]{babel}
\usepackage{graphicx}
\usepackage{epsfig}
\usepackage{bm}
\usepackage{longtable}
\usepackage{verbatim}
\usepackage{longtable}
\usepackage{color}
\usepackage{subfigure}
\usepackage[utf8]{inputenc}
\usepackage{xcolor}
\usepackage{xspace}
\usepackage{ulem}
\usepackage{lscape}
\usepackage{wasysym}
\usepackage[toc,page]{appendix}
\usepackage{mathtools}
\usepackage{multirow,rotating}
\usepackage{stackrel}
\usepackage{cancel}
\usepackage{pifont}
\usepackage{dcolumn}

\newcommand{\mathsym}[1]{{}}

\newcommand{\I}{\ensuremath{\mathrm{i}}}

\renewcommand{\d}{\ensuremath{\mathrm{d}}}

\newcommand{\SU}[1]{\ensuremath{\mathrm{SU}(#1)}}

\renewcommand{\d}{\ensuremath{\mathrm{d}}}

\begin{document}

\title{Spontaneous BRST symmetry breaking in infrared QCD}
\author{Angelo Raffaele Fazio}
\affiliation{Departamento de Física, Universidad Nacional de Colombia, \\ Ciudad Universitaria, Bogot\'a, Colombia}
\email{arfazio@unal.edu.co}
\author{Adam Smetana}
\affiliation{Institute of Experimental and Applied Physics, \\ Czech Technical University in Prague, Prague, Czech Republic}
\email{adam.smetana@cvut.cz}
\date{\today }

\begin{abstract}
We present a novel proposal for the effective Lagrangian of the low-energy Yang--Mills quantum field theory. The proposed effective Lagrangian exhibits the spontaneous BRST symmetry breaking. We built the Fujikawa model that we couple to the Yang--Mills elementary field sector, motivated by the analogy with Chiral Quark Model. We interpret the Fujikawa fields as effective fields composite of the elementary gluon and ghost fields. In order to justify the existence of two massless Nambu--Goldstone modes among the Fujikawa fields, we require not only the BRST but also the anti-BRST invariance of the effective Lagrangian, both to be spontaneously broken. The most striking consequence of that is the emergence of the effective gluon and ghost masses. We reproduce the Curci--Ferrari model as a special case of our effective model upon the spontaneous BRST symmetry breaking. In order to reproduce also the  non-nilpotent modified BRST symmetry, characteristic for the Curci--Ferrari model, we modify our effective Lagrangian to be invariant with respect to the extended-BRST symmetry, which mixes the elementary and Fujikawa field sectors, and which is nilpotent. The Curci--Ferrari is reproduced by the elementary field sector of the resulting Lagrangian. The remaining Fujikawa's field dependent terms guarantee the underlying nilpotent extended-BRST symmetry, which is now hidden in the sense of the spontaneous symmetry breaking.
\end{abstract}

\maketitle
\flushbottom

\tableofcontents

\section{Introduction}

The quantum chromodynamics (QCD) theory of colored fields based on the $\SU{3}_\mathrm{c}$ Yang--Mills theory has proven itself to be a proper perturbative description of hadronic phenomena characterized by an ultraviolet energy transfer. Due to the asymptotic freedom of QCD interaction \cite{Gross1973,Politzer1973}, the harder collision the hadron experiences the more it behaves as a collection of free colored partons, quarks and gluons, and the approximation by perturbative quantum field theory, where colored quarks and gluons are allowed to be asymptotic states, is ever more appropriate. In the infrared regime however quarks and gluons condense into the droplets of colorless hadrons and by the strong QCD interactions they are confined there \cite{Gell-Mann1964,Zweig1964,Wilson1974}. Such a drastic change in number and nature of perturbative degrees of freedom of the theory, resembles phase transitions in condensed matter physics accompanied by the spontaneous symmetry breaking characterized by the nonzero value of some order parameter. 

Due to the increase of its interaction strength in the infrared regime, the QCD nonperturbatively changes its ground state into a perfect color paramagnet \cite{Saito1980}. The chromoelectric field of any colored object placed into such medium is compressed into a flux tube that originates at the object and ends at infinity -- a configuration which is unbearably energetically expensive. Therefore the perturbative particle excitations can only be color neutral composites of quarks and gluons confined together with their chromoelectric fields inside the hadrons and glueballs. 

Since the birth of QCD, many attempts to describe the nontrivial QCD ground state has been made based on, e.g., condensation of the chromomagnetic monopoles in the dual color superconductor picture \cite{tHooft1978}, analyses of the infrared limit of gluon and ghost propagator \cite{Cornwall:1981zr,Alkofer:2000wg,Cucchieri2008,Binosi2026}, ghost condensation \cite{Lemes2002}, etc. Supported by the numerous numerical results of the lattice QCD \cite{Boucaud2001,Bornyakov2021}, these attempts have in common to ascribe the significance to the nonzero expectation value of the colorless composite operator $A^2$ \cite{Zwanziger1989,Gubarev2001,Verschelde2001}, 
\begin{equation}\label{AA}
\langle A^\mu_a A_{\mu a}\rangle\ne0 \,.
\end{equation} 
The interpretation of this vacuum expectation value is not unique and that should not be a surprise as such operator depends on the choice of gauge, thus it is not invariant with respect to the local $\SU{3}_c$ transformation. At the same time, however, it is invariant with respect to the global $\SU{3}_c$ transformation, in other words $\langle A^\mu_a A_{\mu a}\rangle$ does not break any of the eight color group generators represented, in the self- representation, by the Gell-Mann matrices. This should be contrasted with the perturbative regime in ultraviolet, where nonzero $\langle A^\mu_a A_{\mu a}\rangle$ would be in striking conflict with gauge invariance allowing for massless gluon asymptotic states.

In our present work we take these observations literally and consider the phase transition between the confined and deconfined phases of QCD matter being characterized by symmetry breaking pattern
\begin{equation}\label{SU3}
    \left.\SU{3}_c\right|_\mathrm{local} \rightarrow \left.\SU{3}_c\right|_\mathrm{global} \,.
\end{equation}
The operational meaning of this symmetry breaking pattern is however not a priori clear. Namely, it is not apparent how to apply the Nambu--Goldstone theorem when no global symmetry generator appears to be broken. Also it is not clear whether such symmetry breaking is connected with any kind of the Meissner effect by which gluons or other gauge degrees of freedom would acquire their mass by ``eating" the massless Nambu--Goldstone modes. Operationally meaningful and equivalent reformulation of the symmetry breaking pattern \eqref{SU3} is given in terms of a global BRST symmetry,
\begin{equation}
    \left.BRST(\SU{3}_c)\right|_\mathrm{global} \rightarrow \left.\SU{3}_c\right|_\mathrm{global} \,.
\end{equation}
That leads us to the concept of the spontaneous BRST symmetry breaking which has been pioneered by Fujikawa \cite{Fujikawa1983}, and elaborated further by several authors, e.g. \cite{Dudal2012}, ever since then. These works attempt to solve the Gribov problem, which is often in literature related to confinement, by the spontaneous BRST symmetry breaking. However, some argue that soft BRST symmetry breaking is enough, see a review \cite{Vandersickel2012}. These objections are supported by no evidence for the failure of the Kugo--Ojima quartet mechanism \cite{Kugo1979} within the spectrum of the theory, basically pointing towards non-existence of the massless fermionic BRST Nambu--Goldstone mode \cite{Mader2013}, supported also by the lattice analyses \cite{Cucchieri2008}, and similarly by the spectral tests of nonperturbative BRST \cite{Li2021}. On the other hand, if the Gribov problem is taken seriously, the Kugo--Ojima mechanism is not realized trivially in a nonperturbative regime \cite{Burgio2009}. Also the spontaneous BRST symmetry breaking may be a subtle phenomenon of quantum field theory in continuum that might not be grasped by the lattice simulations. The most recent study of emergent Gribov horizon from replica symmetry breaking, as an example, demonstrates that the concept of the spontaneous BRST symmetry breaking in infrared is far from being ruled out. In the present paper we want to contribute to the discussion by following and elaborating further the original work of Fujikawa \cite{Fujikawa1983} in analogy with the chiral quark model for chiral symmetry breaking in QCD \cite{Manohar1984}.

Fujikawa proposed his, so called, non-gauge model of the spontaneous BRST symmetry breaking built in terms of colorless elementary scalar fields forming components of two BRST doublets \cite{Fujikawa1983}. To construct a BRST invariant action he employed the superfield formalism. He proposed a specific model with a potential for the ghost-number-neutral scalar fields designed to have nontrivial BRST symmetry breaking minimum. He then identified the ghost-number-odd scalar fields as the Nambu--Goldstone modes. The Fujikawa model is a prototype model of the spontaneous BRST symmetry breaking, making however no reference and relation to any underlying Yang--Mills gauge theory. On the other hand, some authors consider the spontaneous BRST symmetry breaking directly within a Yang--Mills theory, typically, by extending the Faddeev--Popov sector for additional ghost interactions which may trigger the ghost condensation. 

In the present work we built the Fujikawa model, making the steps which Fujikawa left unperformed. First, we make connection of the Fujikawa fields to the Yang--Mills fields. Very much analogously to the chiral perturbation theory, where the pion effective fields are related to the interpolating composite quark--antiquark operators, we relate the Fujikawa fields to the interpolating operators composed of the elementary colorful (gluon and ghost) fields. This we perform to the lowest order in the operator dimension guided by the requirement that the Nambu--Goldstone modes should inherit the quantum numbers of the broken generators. Second, again in analogy with the chiral perturbation theory, we construct the effective Lagrangian consisting of all possible BRST invariant terms composed of both the elementary colorful fields and the composite colorless fields. Third, doing that, we arrive to the generalization of the Fujikawa model coupled to the elementary field sector of the original gauge Yang--Mills theory. We propose this model as a prototype example of the effective low-energy model of QCD, so far with no presence of quarks, for simplicity. Fourth, under the assumption that the composite scalar potential exhibits a nontrivial minimum, the model describes the spontaneous BRST symmetry breaking. In the analogous way as in the sigma model, the ghost-number-neutral composite scalar fields develop a nonzero vacuum expectation value, creating the mass term of gluons and ghosts, and giving the role of massless Nambu--Goldstone modes to the ghost number odd composite scalar fields. Fifth, we restrict to the most general BRST invariant model by requiring also the anti-BRST invariance as accidental symmetry of the Faddeev--Popov Lagrangian. This requirement justifies the presence of the two Nambu--Goldstone modes, one for the broken BRST generator and the other for the broken anti-BRST generator. This aspect has not been  analyzed by Fujikawa and we obtain a symmetric model under exchange of the two Nambu-Goldstone modes in the sense clarified later. Sixth, it will be proven that our model, upon the spontaneous BRST symmetry breaking, contains the Curci-Ferrari model as its special case given by a specific choice of values of the model parameters. 

\section{Spontaneous BRST symmetry breaking}

We built our model on the Fujikawa model, which we couple to the Yang--Mills elementary field sector, and whose fields, the so called Fujikawa fields, we interpret as effective fields composite of the elementary gluon and ghost fields. In order to justify the existence of two massless Nambu--Goldstone modes among the Fujikawa fields, we require not only the BRST but also the anti-BRST invariance of the effective Lagrangian, both to be spontaneously broken. The most striking consequence of that is the emergence of the effective gluon and ghost masses. These masses are the defining features of the Curci--Ferrari model \cite{Curci1976}. Therefore in this work, we are trying to reproduce the Curci--Ferrari model as a special case of our effective model upon the spontaneous BRST symmetry breaking. As well known, see e.g. \cite{Lavrov2012}, the Curci--Ferrari model exhibits the modified-BRST symmetry which however is not nilpotent. In order to reproduce also the modified-BRST symmetry, we modify our effective Lagrangian to be invariant with respect to the so called extended-BRST symmetry, which mixes the elementary and Fujikawa field sectors, and which is nilpotent. Upon the spontaneous BRST symmetry breaking, the extended-BRST transformation is deformed, in the similar way as shown in \cite{Amaral2022}, to the Curci--Ferrari modified-BRST transformation plus additional terms nonlinear in fields. The additional terms guarantee the underlying nilpotent extended-BRST symmetry, which is now hidden in the sense of the spontaneous symmetry breaking.   

\subsection{Fujikawa non-gauge model}

Fujikawa studied spontaneous breaking of BRST symmetry \cite{Fujikawa1983} motivated by understanding confinement and Gribov problem and the physical meaning of BRST invariance beyond standard gauge theories. One of his key contributions was to show that BRST symmetry can be spontaneously broken even in a model with no gauge redundancy, so called a non-gauge model. He constructed a model with BRST symmetry without gauge symmetry using a quartet of scalar and color singlet fields, denoted there as $\varphi$, $\eta$, $\xi$, and $B$ related by a nilpotent BRST transformation\footnote{In this section we use the same notation as in the Fujikawa's paper. In the remaining of the paper we switch into our notation.}
\begin{subequations}\label{nongaugeBRST}
\begin{eqnarray}
        \delta\varphi &=&\I\theta\eta \,,\ \ \ \delta\eta\,=\,0 \,,\\
        \delta\xi &=&\theta B \,,\ \ \delta B\,=\, 0 \,,
\end{eqnarray}
\end{subequations}
where $\theta$ is a Grassmannian transformation parameter. Fujikawa used an efficient superfield formalism and constructed a BRST invariant model 
\begin{eqnarray}\label{LF0}
    \mathcal{L}^0_\mathrm{F}&=&\tfrac{1}{2}\partial_\mu B\partial^\mu B + \partial_\mu B\partial^\mu\varphi -\I \partial_\mu\xi\partial^\mu\eta - \tfrac{1}{2}M^2 B^2
    -m_0^2(B\varphi-\I\xi\eta)\,,\\
    &&-g(B\varphi-\I\xi\eta)^2-g'B^4-g''B^2(B\varphi-\I\xi\eta)\,.\nonumber
\end{eqnarray}
The potential for the fields $B$ and $\varphi$ develops two degenerate minima with $\iota=\pm1$, 
\begin{eqnarray}
    B_0 &=& -\iota\sqrt{\frac{g'' m_0^2-g M^2}{4g g'-g''^2}} \,,\label{vevB}\\
    \varphi_0 &=& \iota\frac{4g' m_0^2-g'' M^2}{2\sqrt{4g g'-g''^2}\sqrt{g'' m_0^2-g M^2}}. \label{vevphi}
\end{eqnarray}
Upon the shift of the fields $B\rightarrow B+B_0$ and $\varphi\rightarrow \varphi+\varphi_0$ we can write the shifted Lagrangian, here explicitly just in the quadratic part,
\begin{eqnarray}
    \mathcal{L}^0_\mathrm{F}&=&\tfrac{1}{2}\partial_\mu B\partial^\mu B + \partial_\mu B\partial^\mu\varphi -\I \partial_\mu\xi\partial^\mu\eta - (B\ \ \varphi)\mathcal{M}\left(\begin{array}{c} B \\ \varphi \end{array}\right)+\dots\,,
\end{eqnarray}
where
\begin{eqnarray}
    \mathcal{M}&=&\left(\begin{array}{cc} M^2/2+g\varphi_0^2+6g' B_0^2+3g''B_0\varphi_0 & m^2+4gB_0\varphi_0+3g''B_0^2 \\ m^2+4gB_0\varphi_0+3g''B_0^2 & gB_0^2 \end{array}\right)\,.
\end{eqnarray}
The quadratic Lagrangian of $B$ and $\varphi$ must be diagonalized and it leads to two massive fields that are linear combinations of $B$ and $\varphi$. On the other hand $\xi$ and $\eta$ are massless fields that represent Nambu--Goldstone modes of the spontaneously broken BRST symmetry \eqref{nongaugeBRST}. 

In \cite{Fujikawa1983} Fujikawa addressed a close relation of the dynamical stability of the BRST symmetry to the Gribov problem of the Yang--Mills theory, by pointing out that the potential existence of the degenerate vacua is related to the degenerate gauge field configurations. He considered $\langle0|b^a b^a|0\rangle\ne0$ as the order parameter of the BRST symmetry breaking being connected to the composite operator $b^a\bar{c}^a$ that contains the Grassmannian massless Nambu--Goldstone boson as it follows from
\begin{equation}\label{FNGcommutator}
    \langle0|\{Q,b^a\bar{c}^a\}|0\rangle= \langle0|b^a b^a|0\rangle\ne0 \,.
\end{equation}

\subsection{General considerations}

Applying the Nambu--Goldstone theorem to the case of the spontaneously broken BRST charge $Q$ consists of identifying the operator that creates the Nambu--Goldstone mode. We denote such Nambu--Goldstone operator as $\bar\pi$. Direct consequence of the Nambu--Goldstone theorem is that the Nambu--Goldstone modes inherit the quantum numbers from the broken generators. The general condition for the existence of the Nambu--Goldstone massless mode is
\begin{equation}\label{NGcommutator}
    \langle0|\{Q,\bar\pi\}|0\rangle= \langle\bar\phi\rangle\ne0 \,,
\end{equation}
whose one possible realization is \eqref{FNGcommutator}. The right hand side, $\langle\bar\phi\rangle$, is a general order parameter of the spontaneously broken BRST symmetry given as the vacuum expectation value of some operator $\bar\phi$. In our case, $\bar\phi$ should be a Lorentz scalar and a color singlet, not to violate the Lorentz invariance or the global $\SU{3}_\mathrm{c}$. Because $Q$ is an anticommuting operator, Nambu--Goldstone field $\bar\pi$ must be an anticommuting field, that's why on the left-hand side there is the anticommutator. 

Here, however, we see peculiarity of the spontaneous BRST symmetry breaking: the anticommutator actually represents the BRST transformation of the Nambu--Goldstone field, $\delta\bar\pi=\{Q,\bar\pi\}\equiv\bar\phi$. Hence the field $\bar\phi$ on the right hand side is a BRST invariant, because of nilpotency of the BRST, as $\delta\bar\phi=\delta\delta\bar\pi=0$. Yet the assumption of $\langle\bar\phi\rangle\ne0$ implies the spontaneous BRST symmetry breaking as it requires 
\begin{equation}
    Q|0\rangle\ne0.
\end{equation}

On the other hand, we infer that if the BRST is spontaneously broken, there should exist a BRST non-invariant operator, denoted here as $\varphi$ with $\delta\varphi\ne0$, whose vacuum expectation value is nonzero 
\begin{equation}
\langle\varphi\rangle\ne0 \,.
\end{equation}
Again, it should be Lorentz scalar and a color singlet. Obviously, it cannot be any BRST exact object such as $\bar{\phi}$, nor $b^ab^a$, but it can be $A^\mu_a A_{\mu a}$. Let us denote the BRST transform of $\varphi$ by $\pi$ so that 
\begin{equation}
    \delta\varphi=\theta\pi\,,\ \ \delta\pi=0\,.
\end{equation}
That, actually, completes the quartet of the Fujikawa fields identifying $\{\varphi,\eta,\xi,B\}\rightarrow\{\varphi,\pi,\bar\pi,\bar\phi\}$. 

If, according to the Fujikawa construction, $\pi$ is to be the second Nambu--Goldstone mode, then there should exist some other broken generator, whose anticommutator has a nonvanishing vacuum expectation value, in the same line as \eqref{NGcommutator}. Such generator exists and it is the generator of the anti-BRST transformation $\bar{Q}$, so that
\begin{equation}
    \langle0|\{\bar{Q},\pi\}|0\rangle= \langle\phi\rangle\ne0 \,,
\end{equation}
where $\phi=-\bar\phi$ by the requirement of the nilpotency of the anti-BRST transformation, see Appendix \ref{A1}.

We conclude this section by emphasizing that the Fujikawa model with its BRST--anti-BRST quartet field content is not just a simplistic toy model, but it provides a necessary Higgs--Nambu--Goldstone field sector of any quantum field theory that exhibits the spontaneous BRST symmetry breaking, such that, potentially, the Yang--Mills theory of QCD.

\section{Effective composite fields}

In order to effectively describe the spontaneous BRST symmetry breaking we extend the field content of the Yang--Mills theory for the Fujikawa fields $\bar\pi$, $\bar\phi$, $\varphi$ and $\pi$ as the effective field being composites of the elementary gluon and ghost fields. 

\subsection{The first Fujikawa BRST doublet}
We start by introducing the mass dimension $+1$ and ghost number $-1$ field $\bar\pi$. Based on its required properties we can identify its possible lowest-dimensional interpolating operators denoted by the same symbol but with an index:
\begin{subequations}\label{barpi}
\begin{eqnarray}
    \bar\pi\quad\propto\quad
    \bar\pi_1 &\equiv& \bar{c}\cdot(A\otimes A) \,,\\
    \bar\pi_2 &\equiv& \bar{c}\cdot(\bar{c}\times c) \,,\\
    \bar\pi_3 &\equiv& A\cdot\partial \bar{c} \,,\\
    \bar\pi_4 &\equiv& \bar{c}\cdot b \,,\\
    \bar\pi_5 &\equiv& \bar{c}\cdot \partial A \,,
\end{eqnarray}
\end{subequations}
where we use a compact notation of suppression of contracted Lorentz indices and 
\begin{equation}
x\cdot y\equiv x_ay_a\,,\quad (y\times z)_a\equiv f_{abc}y_bz_c\,,\quad(y\otimes z)_a\equiv d_{abc}y_bz_c\,.
\end{equation}
The anticommutator in \eqref{NGcommutator} represents the BRST transformation of the field $\bar\pi$ so that 
\begin{eqnarray}
    \delta\bar\pi &=& \theta\bar\phi\,,
\end{eqnarray}
where $\theta$ is a Grassmanian BRST transformation parameter of the mass dimension $-1$, therefore we end up with the dimension $+2$ for the effective field $\bar\phi$. The interpolating operators for $\bar\phi$ are given by the BRST transformation of the interpolating operators $\bar\pi_i$, $i=1,\dots,5$, and are denoted as $\bar\phi_i$ 
\begin{equation}\label{Qbarpi}
    \{Q,\bar\pi_i\}=\bar\phi_i \,.
\end{equation}
It is useful to write their explicit forms
\begin{subequations}\label{barphi}
\begin{eqnarray}
    \bar\phi\quad\propto\quad
    \bar{\phi}_1 &=& b\cdot(A\otimes A) - 2A\cdot(\bar{c}\otimes Dc) \,,\\
    \bar{\phi}_2 &=& 2b\cdot(\bar{c}\times c) +g(\bar{c}\times c)\cdot(\bar{c}\times c) \,,\\
    \bar{\phi}_3 &=& A\cdot\partial b -\partial\bar{c}\cdot Dc \,,\\
    \bar{\phi}_4 &=& b\cdot b \,,\\
    \bar{\phi}_5 &=& b\cdot\partial A -\bar{c}\cdot\partial Dc\,.
\end{eqnarray}
\end{subequations}
By the nilpotency of the BRST transformation for the composite operators  we have 
\begin{eqnarray}
    \delta\bar\phi_i &=& 0\,,
\end{eqnarray}
and we require the same for the effective field 
\begin{eqnarray}
    \delta\bar\phi &=& 0\,.
\end{eqnarray}
These transformation properties are enough to construct a BRST invariant effective Lagrangian simply as the most general functional of the BRST invariants $\bar\phi$ and $\bar\phi_i$, and their derivatives. But such model in no way can play a role of the model of the spontaneous BRST symmetry breaking for two reasons. At first place, any shift of the field $\bar\phi$ by its nonzero vacuum expectation value $\langle\bar\phi\rangle$ gives another effective Lagrangian, which is however still a functional $\bar\phi$ and $\bar\phi_i$ and their derivatives. Second, the Nambu--Goldstone field $\bar\pi$ is decoupled. Therefore in the next subsection we introduce the second Fujikawa BRST doublet.

\subsection{The second Fujikawa BRST doublet}

In \eqref{barpi} we have introduced the basis for the interpolating operators for the Nambu--Goldstone field $\bar\pi$. Analogously, we can introduce the basis for the interpolating operators for the second Nambu--Goldstone field $\pi$ just by interchanging the ghost and antighost fields
\begin{subequations}\label{pi}
\begin{eqnarray}
    \pi\quad\propto\quad
    \pi_1 &\equiv& c\cdot(A\otimes A) \,,\\
    \pi_2 &\equiv& c\cdot(c\times\bar{c}) \,,\\
    \pi_3 &\equiv& A\cdot\partial c \,,\\
    \pi_4 &\equiv& c\cdot b \,,\\
    \pi_5 &\equiv& c\cdot \partial A \,.
\end{eqnarray}
\end{subequations}
In order to complete the Fujikawa quartet we need to find the possible interpolating operators for the field $\varphi$ whose BRST transformation gives $\pi$:
\begin{equation}
    \delta\varphi=\theta\pi \,.
\end{equation}
Of the lowest mass dimensionality, $+2$, there are, actually, just two of them
\begin{subequations}\label{varphi}
\begin{eqnarray}
    \varphi\quad\propto\quad
    \varphi_2 &\equiv& \bar{c}\cdot c \,,\\
    \varphi_3 &\equiv& A\cdot A \,.
\end{eqnarray}
\end{subequations}
Their BRST transformation gives only specific linear combinations of the operators $\pi_i$ 
\begin{eqnarray}
    {}[Q,\varphi_2] &=& \pi_4+ \frac{g}{2}\pi_2\,,\\
    {}[Q,\varphi_3] &=& 2\pi_3\,.
\end{eqnarray}
Therefore only these can play the role of the interpolating operators for the Nambu--Goldstone field $\pi$.

\subsection{The Fujikawa BRST and anti-BRST quartet}

As we have already stressed out, to justify the existence of both Nambu--Goldstone modes, $\bar\pi$ and $\pi$, we should consider also the anti-BRST symmetry on top of the BRST symmetry that are spontaneously broken. The Fujikawa fields then form a BRST--anti-BRST quartet of fields satisfying
\begin{eqnarray}
    &&\{\bar{Q},[Q,\varphi]\}=\{\bar{Q},\pi\}=\phi=-\bar\phi \,,\\
    &&\{Q,[\bar{Q},\varphi]\}=\{Q,\bar\pi\}=\bar\phi=-\phi \,.
\end{eqnarray}
From this algebraic structure it follows that the interpolating operators for the Fujikawa quartet effective fields are given by a two-parametric linear combination:
\begin{eqnarray}
    \varphi&\sim&x\,\varphi_2+y\,\varphi_3 \,,\\
    \bar\pi&\sim&x\left(\bar\pi_4+ \frac{g}{2}\bar\pi_2\right)+2 y\bar\pi_3 \,,\\
    \pi&\sim&x\left(\pi_4+ \frac{g}{2}\pi_2\right)+2 y\pi_3 \,,\\
    \bar\phi=-\phi&\sim&x\left(\bar{\phi}_4+ \frac{g}{2}\bar{\phi}_2\right)+2 y\bar{\phi}_3 \,,
\end{eqnarray}
where $x$ and $y$ are of mass dimension $-2$ phenomenological parameters not specified at this level.

\section{The effective Lagrangian}

We formulate the effective model of the low-energy QCD, omitting the quark sector in this work, by specifying the effective Lagrangian $\mathcal{L}_\mathrm{eff}$: 
\begin{eqnarray}\label{eL}
    \mathcal{L}_\mathrm{eff}=-\frac{1}{4}F\cdot F+\mathcal{L}_\mathrm{g}+\mathcal{L}_\mathrm{F}+\mathcal{L}_\mathrm{gF}\,.
\end{eqnarray}
We supplement the first term, which is the classical QCD gauge invariant gluon Lagrangian, by the ghost sector Lagrangian $\mathcal{L}_\mathrm{g}$ in order to fix the gauge, and innovatively by the effective Fujikawa field sector Lagrangian $\mathcal{L}_\mathrm{F}$ containing purely Fujikawa field terms and $\mathcal{L}_\mathrm{gF}$ containing terms that mix the effective Fujikawa fields with elementary ghost and gluon fields. This approach follows the same spirit as the Chiral Quark Model \cite{Manohar1984}, which addresses the chiral symmetry breaking of QCD by introducing pion fields - the Nambu--Goldstone bosons - within the nonlinear representation into the quark Lagrangian.

We construct these terms to be BRST invariant within the framework of the superfield formalism described in Appendix~\ref{App_Superfields}. We use the superfields ${\cal{A}}^a_\mu(x,\theta)$, ${\cal{C}}^a(x,\theta)$ and $\bar{\cal{C}}^a(x,\theta)$ that contain the elementary ghost and gluon fields, and the superfields $\xi(x,\theta)$ and $\varphi(x,\theta)$ that contain the effective Fujikawa fields including the BRST Nambu--Goldstone fields. By the general logic of the spontaneous symmetry breaking the BRST ``pions" should be embedded in the BRST non-linear representations, that are exactly the superfields $\xi(x,\theta)$ and $\varphi(x,\theta)$. 

In general, the BRST invariant Lagrangian is formed by two types of terms. One is the gauge invariant part, the BRST cohomology, whose example is the first term in \eqref{eL}, the gauge invariant gluon Lagrangian. The other one is the BRST exact part that can be written as
\begin{eqnarray}
    \int\mathrm{d}x\mathcal{L}_\mathrm{exact}&=&\int\mathrm{d}x\int\mathrm{d}\theta\Psi(x,\theta)\,,
\end{eqnarray}
where $\Psi(x,\theta)$ is the Lorentz-scalar superfield in color-singlet and dimension $-1$. In order to guarantee the power-counting renormalizability $\mathrm{dim}\Psi\leq3$. This is the way we construct $\mathcal{L}_\mathrm{g}$, $\mathcal{L}_\mathrm{F}$ and $\mathcal{L}_\mathrm{gF}$ in the following subsections. 

As we discussed above, the anti-BRST invariance of the model is essential for realizing spontaneous BRST symmetry breaking, in order to justify the existence of both Fujikawa Nambu--Goldstone modes. Therefore, at the end of this section, we apply an additional constraint on the BRST invariant effective Lagrangian, that of invariance with respect to the anti-BRST transformation $\bar\delta$, requiring
\begin{equation}
    \bar\delta\int\mathrm{d}x\mathcal{L}_\mathrm{eff}=0\,.
\end{equation}

\subsection{The ghost sector Lagrangian}

By means of the superfield construction, the ghost Lagrangian $\mathcal{L}_\mathrm{g}$ up to the renormalizable level, i.e. $\mathrm{dim}\leq4$, is given by the superfield $\Psi(x,\theta)$ consisting of linear combination of the composite operators $\xi_i(x,\theta)$:
\begin{eqnarray}\label{Lgf}
    \int\mathrm{d}x\mathcal{L}_\mathrm{g}&=&\int\mathrm{d}x\int\mathrm{d}\theta\sum_{i=1,\dots,5}\kappa_i\xi_i(x,\theta)=\int\mathrm{d}x\sum_{i=1,\dots,5}\kappa_i\bar\phi_i\,\\
    &=&\int\mathrm{d}x\Big\{ (\kappa_3-\kappa_5)(A\cdot\partial_\mu b_a-\partial\bar{c}\cdot Dc)+\kappa_4b\cdot b + \kappa_2 \Big[2b\cdot(\bar{c}\times c) +g(\bar{c}\times c)\cdot(\bar{c}\times c)\Big ] \\
    && \quad\quad\quad + \kappa_1 \Big[b\cdot(A\otimes A) - 2A\cdot(\bar{c}\otimes Dc)\Big] \Big\}.\nonumber
\end{eqnarray}
Clearly, the normalization of the ghost kinetic term requires
\begin{equation}
    (\kappa_3-\kappa_5)=1\,.
\end{equation}
Without loss of generality, we can set $\kappa_5=0$ because by integration by parts $\bar\phi_5$ does not provide any input to the action other than that of $\bar\phi_3$. 

One of the special cases of the Lagrangian \eqref{Lgf} is of course the Faddeev--Popov Lagrangian with the parameter choice $\kappa_1=0$, $\kappa_2=0$, $\kappa_3=1$, and $\kappa_4=1/2\alpha$. Another special case is the Lagrangian of the Curci--Ferrari model \cite{Curci1976} in massless limit with the parameter choice $\kappa_1=0$, $\kappa_2=\I g(\beta-1)/4\alpha$, $\kappa_3=1$, and $\kappa_4=1/2\alpha$.

\subsection{The generalized Fujikawa Lagrangian}

The third term of \eqref{eL} encounters for the realization of the, so called, non-gauge model of Fujikawa \eqref{LF0} (hence the subscript $\mathrm{F}$). Equipped with the superfields \eqref{Ff} we can construct the BRST invariant Lagrangian as a sum of operators constructed as the $\theta$-integral of the mass dimension $-1$ superfield. The general non-derivative BRST invariant operator then is written as
\begin{equation}
    \label{Vmn}\int\d\theta\,\big[\xi\varphi^n(\partial_\theta\xi)^{m}\big]\equiv \mathcal{V}^{mn}= \bar\phi^{m}(\bar\phi\varphi - n\bar\pi\pi)\varphi^{n-1} \,,\ \mathrm{for}\ m,n=0,1,2,\dots\,. \\
\end{equation}

Due to anticommutativity, all operators with second and higher power of the superfield $\xi$ is zero. Due to anticommutativity of $\pi(x)$ and $\bar\pi(x)$, together with the ghost number conservation, the fields $\pi(x)$ and $\bar\pi(x)$ appear in the BRST invariant operators only in the combination $\bar\pi\pi$ and in its first power. It also means that the bilinear $\bar\pi\pi$ cannot acquire any vacuum expectation value.

Consequently we can write the general potential for the composite fields
\begin{equation}\label{V}
    \mathcal{V}=\sum_{m,n}a_{mn}\mathcal{V}^{mn}=v^2\bar\phi f[\tfrac{\bar\phi}{v^2},\varphi-\tfrac{\bar\pi\pi}{\bar\phi}\big]\,,
\end{equation}
where $f$ is a regular function of two variables that can be expanded into series with coefficients $a_{mn}$.

The general derivative terms are constructed in the analogy with \eqref{Vmn} equipped with even number of derivatives with mutually contracted Lorentz indices. The lowest mass dimensional terms with two derivatives form the kinetic terms for the Fujikawa fields
\begin{eqnarray}
     \textstyle{\int}\d\theta\,\big[\partial_\mu\xi\partial^\mu\varphi\big]&=& \partial_\mu\bar\phi\partial^\mu\varphi - \partial_\mu\bar\pi\partial^\mu\pi \,, \\
     \textstyle{\int}\d\theta\,\big[\partial_\mu\xi\partial^\mu(\partial_\theta\xi)\big]&=& \partial_\mu\bar\phi\partial^\mu\bar\phi \,.
\end{eqnarray}

Putting all together we can write the generalized Fujikawa Lagrangian as 
\begin{eqnarray}\label{LFx}
    \mathcal{L}_\mathrm{F}&=&\tfrac{1}{2v^2}\partial_\mu\bar\phi\partial^\mu\bar\phi + \partial_\mu\bar\phi\partial^\mu\varphi - \partial_\mu\bar\pi\partial^\mu\pi 
   - v^2\bar\phi f\big[\tfrac{\bar\phi}{v^2},\varphi-\tfrac{\bar\pi\pi}{\bar\phi}\big]\,,
\end{eqnarray}
where we have introduced the dimensionful parameter $v$ in order to account with dimensionality of the Lagrangian. 
Upon the field re-normalization $\bar\phi\leftarrow\bar\phi/v$ and $\varphi\leftarrow\varphi\,v$ we can write the renormalizable $\mathrm{dim}\leq4$ part
\begin{eqnarray}\label{LF}
    \mathcal{L}_\mathrm{F}&=&\tfrac{1}{2}\partial_\mu\bar\phi\partial^\mu\bar\phi + \partial_\mu\bar\phi\partial^\mu\varphi - \partial_\mu\bar\pi\partial^\mu\pi - v^3 a_{00}\, \bar\phi - v^2 a_{10}\, \bar\phi^2 - v a_{20}\, \bar\phi^3 - a_{30}\, \bar\phi^4 \\
   &&
   \hspace{4.75cm}-\, v^2 a_{01} (\bar\phi\varphi-\bar\pi\pi) - v a_{11} \bar\phi(\bar\phi\varphi-\bar\pi\pi) - a_{21} \bar\phi^2(\bar\phi\varphi-\bar\pi\pi) \nonumber\\
   &&
   \hspace{4.75cm}-\, v a_{02} \varphi(\bar\phi\varphi-2\bar\pi\pi) - a_{12} (\bar\phi\varphi-\bar\pi\pi)^2 - a_{03} \varphi^2(\bar\phi\varphi-3\bar\pi\pi) \,.\nonumber
\end{eqnarray}
The Fujikawa model \eqref{LF0} is obtained from \eqref{LF} by $a_{00}=0$,  $a_{10}=\tfrac{M^2}{2v^2}$, $a_{20}=0$, $a_{30}=g'$, $a_{01}=\tfrac{m_0^2}{v^2}$, $a_{11}=0$, $a_{21}=g''$, $a_{02}=0$, $a_{12}=g$, and $a_{03}=0$. In that case, the potential $\mathcal{V}$ is minimized by the vacuum expectation values \eqref{vevB} and \eqref{vevphi}. 

The general potential $\mathcal{V}$ such as in \eqref{LF} is minimized by $\langle\bar\phi\rangle$ and $\langle\varphi\rangle$ of more general form. If these are nontrivial 
\begin{eqnarray}
    \langle\varphi\rangle&\ne& 0 \,, \\
    \langle\bar\phi\rangle&\ne& 0 \,,    
\end{eqnarray}
then the spontaneous BRST symmetry breaking happens. They can be such that $\langle\varphi\rangle$ vanishes but $\langle\bar\phi\rangle$ should not. It is the necessary condition for the spontaneous BRST symmetry breaking that $\langle\bar\phi\rangle\ne0$. Upon the shift of the fields $\bar\phi\rightarrow\bar\phi+\langle\bar\phi\rangle$ and $\varphi\rightarrow\varphi+\langle\varphi\rangle$ we recover the masslessness of $\pi$ and $\bar\pi$ being the Nambu--Goldstone fields.

%
%
%
By the field redefinition
\begin{equation}\label{shift_phi}
    \bar\phi\longrightarrow\bar\phi-v^2\varphi\,,
\end{equation}
the kinetic term for $\bar\phi$ and $\varphi$ in \eqref{LFx} can be diagonalized
\begin{eqnarray}
    \mathcal{L}_\mathrm{F}&=&\tfrac{1}{2v^2}\partial_\mu\bar\phi\partial^\mu\bar\phi
    - \tfrac{v^2}{2}\partial_\mu\varphi\partial^\mu\varphi - \partial_\mu\bar\pi\partial^\mu\pi  - \mathcal{V}
\end{eqnarray}
The wrong sign of the $\varphi$ and $\pi$ kinetic terms signal their ghost nature.

\subsection{Colorful--Colorless Portal Lagrangian}

The fourth term in \eqref{eL} mixes the elementary colorful and composite (effective) colorless (super)fields. In order to get the renormalizable Lagrangian, we use all possible operators of $\mathrm{dim}\leq4$. Hence we use the superfield operators of dimension not larger than three that are composed as products of superfields $\varphi(x,\theta)$ and $\xi(x,\theta)$, and the composite superfield operators $\varphi_j(x,\theta)$ and $\xi_i(x,\theta)$.
\begin{eqnarray}
    \int\d x\, \mathcal{L}_\mathrm{gF}&=& \int\d x\int\d\theta\, \Big[ \sum_{j=2,3}k_j\varphi_j\xi + \sum_{i=1,\dots,5}\tilde\kappa_i\xi_i\varphi\Big]  \,\\
    &=& \int\d x\, \Big[ k_3(\varphi_3\bar\phi + 2\pi_3\bar\pi)+k_2(\varphi_2\bar\phi + \pi_4\bar\pi +\frac{g}{2}\pi_2\bar\pi) + \sum_{i=1,\dots,5}\tilde\kappa_i(\bar\phi_i\varphi-\bar\pi_i\pi) \Big]  \,
\end{eqnarray}
where $k_j$ and $\tilde\kappa_i$ are phenomenological effective dimensionless coupling parameters.


\subsection{BRST and anti-BRST invariant effective Lagrangian}

So far we have constructed the most general BRST Lagrangian using the superfield formalism. We expect that requirement of its anti-BRST invariance further constrains the number of free parameters. We proceed by checking the anti-BRST invariance of the terms of the general Lagrangian.

First, the Lagrangian $\mathcal{L}_\mathrm{g}$ is anti-BRST invariant simply because
\begin{equation}
    \bar\delta\int\mathrm{d}\theta\xi_i=\bar\delta\bar\phi_i=0\,.
\end{equation}
This is just another realization of the accidental anti-BRST invariance analogous to the case of the Faddeev--Popov Lagrangian. Second, the Lagrangian $\mathcal{L}_\mathrm{F}$ is anti-BRST invariant because
\begin{eqnarray}
    \bar\delta\bar\phi^n &=& 0 \,,\\
    \bar\delta(\bar\phi\varphi^m-m\bar\pi\pi\varphi^{m-1}) &=& 0 \,.
\end{eqnarray}

Third, only the Lagrangian $\mathcal{L}_\mathrm{gF}$ encounters  some level of reduction in order to become anti-BRST invariant. To see that we first analyze the anti-BRST transformation  of the Lagrangian
\begin{eqnarray}
\bar\delta\int\mathrm{d}\theta\big[k_2\varphi_2\xi+k_3\varphi_3\xi\big]&=&\bar\delta\big\{k_2\big[\varphi_2\bar\phi+(\pi_4+\tfrac{g}{2}\pi_2)\bar\pi\big]+k_3\big(\varphi_3\bar\phi+2\pi_3\bar\pi\big)\big\} \\
&=&\bar\theta\big\{k_2\big[(\bar\pi_4+\tfrac{g}{2}\bar\pi_2)\bar\phi-(\bar\phi_4+\tfrac{g}{2}\bar\phi_2)\bar\pi\big]+2k_3\big(\bar\pi_3\bar\phi-\bar\phi_3\bar\pi\big)\big\} \,,\nonumber\\
\bar\delta\int\mathrm{d}\theta\tilde{\kappa}_i\varphi\xi_i&=&\bar\delta\tilde{\kappa}_i\big(\varphi\bar\phi_i+\pi\bar\pi_i\big) \\
&=&\bar\theta\tilde{\kappa}_i\big(\bar\pi\bar\phi_i-\bar\phi\bar\pi_i+\pi\bar\eta_i\big) \,.\nonumber
\end{eqnarray}
Both types of Lagrangian terms are needed in order to provide the anti-BRST invariance. The results of the anti-BRST transformation cancel out if 
\begin{equation}
    \tilde{\kappa}_1=0\,,\ \ \tilde{\kappa}_2=\tfrac{g}{2}k_2\,,\ \ \tilde{\kappa}_3=2k_3\,,\ \ \tilde{\kappa}_4=k_2\,,\ \ \tilde{\kappa}_5=0\,.
\end{equation}
It is just matter of consistency that the last term proportional to $\kappa_i\pi\bar\eta_i$ cancels out as well:
\begin{equation}
    (\bar\eta_4+\tfrac{g}{2}\bar\eta_2)k_2=0\,.
\end{equation}

All in all, the BRST and anti-BRST invariant model is given by the Lagrangian
\begin{eqnarray}
    \mathcal{L}_\mathrm{eff} &=& -\tfrac{1}{4}F\cdot F+\mathcal{L}_\mathrm{F}+\int\mathrm{d}\theta\Big[{\textstyle\sum}\kappa_i\xi_i+k_2(\varphi_2\xi+\varphi\xi_4+\tfrac{g}{2}\varphi\xi_2)+k_3(\varphi_3\xi+2\varphi\xi_3)\Big] \\
    &=& -\tfrac{1}{4}F\cdot F+\mathcal{L}_\mathrm{F}+\Big\{{\textstyle\sum}\kappa_i\bar\phi_i+(k_2\varphi_2+k_3\varphi_3)\bar\phi+\varphi\big[k_2(\bar\phi_4+\tfrac{g}{2}\bar\phi_2)+2 k_3\bar\phi_3)\big] \\
    && \quad\quad\quad\quad\quad\quad\quad\quad+\big[k_2(\pi_4+\tfrac{g}{2}\pi_2)+2k_3\pi_3)\big]\bar\pi+\pi\big[k_2(\bar\pi_4+\tfrac{g}{2}\bar\pi_2)+2k_3\bar\pi_3)\big]\Big\} \nonumber\\
    &=& -\tfrac{1}{4}F\cdot F +\mathcal{L}_\mathrm{F} + \label{Leff}\\
    && + (\kappa_3-\kappa_5) \big[A\cdot\partial b -\partial\bar{c}\cdot Dc\big] + \kappa_4 b\cdot b + \kappa_2\big[2b + g(\bar{c}\times c)\big]\cdot(\bar{c}\times c)  +\kappa_1 \big[b\cdot(A\otimes A) - 2A\cdot(\bar{c}\otimes Dc)\big] \nonumber\\
    &&+\big[k_3 A\cdot A + k_2\bar{c}\cdot c\big]\bar\phi + \big[2 k_3(A\cdot\partial b -\partial\bar{c}\cdot Dc) + k_2 b\cdot b  +\tfrac{g}{2}k_2\big(2b + g\bar{c}\times c\big)\cdot \bar{c}\times c\big]\varphi  \nonumber\\
    &&+ \big[2k_3 A\cdot\partial c + k_2\big(c\cdot b+\tfrac{g}{2}c\cdot\bar{c}\times c\big)\big]\bar\pi - \big[2k_3 A\cdot\partial \bar{c} + k_2\big(\bar{c}\cdot b+\tfrac{g}{2}\bar{c}\cdot\bar{c}\times c\big)\big]\pi \nonumber
\end{eqnarray}

\subsection{General BRST and anti-BRST invariant effective Lagrangian}

Here we closely follow the same line of reasoning as that of the Chiral Quark Model \cite{Manohar1984}, where the nonlinear representation containing pion fields in arbitrarily high power, is coupled to the elementary quark fields that are kept only in renormalizable operators, i.e., only up to dimension four operators. 
Following that we come to the more general effective Lagrangian that differs from \eqref{Leff} by making the $\kappa$ and $k$ parameters depending on the Fujikawa fields, hence $\kappa\rightarrow\mathcal{K}(\varphi,\bar\phi)$ and $k\rightarrow K(\varphi,\bar\phi)$. 

In the superfield formalism the general gluon-ghost-Fujikawa interaction terms are 
\begin{eqnarray}
    \mathcal{L}^\mathrm{gen}_\mathrm{gF} 
    = \sum_{i=1,\dots,5}A^{(i)}_{nm}\int\d\theta\, \xi_i\varphi^n(\partial_\theta\xi)^m 
    + \sum_{i=2,3}B^{(i)}_{nm}\int\d\theta\, \xi\varphi_i\varphi^{n-1}(\partial_\theta\xi)^m 
    + \sum_{i=1,\dots,5}C^{(i)}_{nm}\int\d\theta\, \xi\varphi^n\partial_\theta\xi_i(\partial_\theta\xi)^{m-1}.
\end{eqnarray}
The requirement of the simultaneous anti-BRST invariance,
\begin{eqnarray}
    \bar\delta\int\mathrm{d}x\mathcal{L}^\mathrm{gen}_\mathrm{gF}&=&0\,,
\end{eqnarray}
leads to 
\begin{eqnarray}
    A^{(1)}_{nm} =0 \,,\ \ A^{(2)}_{nm} =\frac{g}{2}\frac{1}{n}B^{(2)}_{nm} \,,\ \ A^{(3)}_{nm} =\frac{1}{n}B^{(3)}_{nm} \,,\ \ A^{(4)}_{nm} =\frac{1}{n}B^{(2)}_{nm} \,,\ \ A^{(5)}_{nm} =0 \,,
\end{eqnarray}
in terms of the coefficient of the functionals
\begin{eqnarray}
    K_{2,3}(\varphi,\bar\phi) &=& \sum_{nm}A^{(2,3)}_{nm}\varphi^n\bar\phi^m \,,\\
    \mathcal{K}_{i=1,\dots,5}(\varphi,\bar\phi) &=& \sum_{nm}C^{(i)}_{nm}\varphi^n\bar\phi^m \,.
\end{eqnarray}
The resulting general effective Lagrangian is
\begin{eqnarray}
    \mathcal{L}^\mathrm{gen}_\mathrm{eff} &=&  -\tfrac{1}{4}F\cdot F +\mathcal{L}_\mathrm{F} + {\textstyle\sum}\mathcal{K}_i\left(\varphi-\tfrac{\bar\pi\pi}{\bar\phi},\bar\phi\right)\bar\phi_i \label{Leff_gen}\\
    && +\big[K_3'(\varphi-\tfrac{\bar\pi\pi}{\bar\phi},\bar\phi)\varphi_3 + K_2'(\varphi-\tfrac{\bar\pi\pi}{\bar\phi},\bar\phi)\varphi_2\big]\bar\phi + \big[2 K_3(\varphi,\bar\phi)\bar\phi_3 + K_2(\varphi,\bar\phi) \big(\bar\phi_4  +\tfrac{g}{2}\bar\phi_2\big)\big]  \nonumber\\
    &&+ \big[2K_3'(\varphi,\bar\phi) \pi_3 + K_2'(\varphi,\bar\phi)\big(\pi_4+\tfrac{g}{2}\pi_2\big)\big]\bar\pi - \big[2K_3'(\varphi,\bar\phi)\bar\pi_3 + K_2'(\varphi,\bar\phi)\big(\bar\pi_4+\tfrac{g}{2}\bar\pi_2\big)\big]\pi \nonumber
\end{eqnarray}
where the prime means the partial derivative with respect to the first argument, $K_{2,3}'(x,y)\equiv\frac{\partial K_{2,3}(x,y)}{\partial x}$. The previous effective Lagrangian \eqref{Leff} is obtained by choosing functions $K_{2,3}(\varphi,\bar\phi)$ to be just linear functions of $\varphi$ and independent on $\bar\phi$: 
\begin{eqnarray}
    K_{2,3}(\varphi,\bar\phi) = k_{2,3} \varphi \,.
\end{eqnarray}

\section{Spontaneous BRST symmetry breaking in the effective model}

Depending on the shape of the Fujikawa potential $\mathcal{V}$ \eqref{V}, the Fujikawa scalar fields $\varphi$ and $\bar\phi$ develop the nontrivial vacuum expectation values and the spontaneous BRST symmetry breaking happens. Shifting the fields
\begin{subequations}\label{SSBshift}
\begin{eqnarray}
    \bar\phi &\rightarrow& \bar\phi+\langle\bar\phi\rangle \,,\\
    \varphi &\rightarrow& \varphi+\langle\varphi\rangle \,,
\end{eqnarray}    
\end{subequations}
in the effective Lagrangian $\mathcal{L}_\mathrm{eff}$ \eqref{Leff}, both gluon and ghost field acquire their effective mass parameters:
\begin{eqnarray}
    M_A &=& k_3\langle\bar\phi\rangle \,,\\
    M_c &=& k_2\langle\bar\phi\rangle \,,
\end{eqnarray}
or from the general effective Lagrangian $\mathcal{L}^\mathrm{gen}_\mathrm{eff}$ \eqref{Leff_gen} the masses get the form:
\begin{eqnarray}
    M_A &=& K_3'(\langle\varphi\rangle,\langle\bar\phi\rangle)\langle\bar\phi\rangle \,,\\
    M_c &=& K_2'(\langle\varphi\rangle,\langle\bar\phi\rangle)\langle\bar\phi\rangle \,.
\end{eqnarray}

\subsection{Derivation of the Curci--Ferrari Lagrangian}
This observation actually reminds strongly the Curci--Ferrari model, whose Lagrangian is
\begin{eqnarray}\label{LCF}
    \mathcal{L}_\mathrm{CF} &=&  -\tfrac{1}{4}F\cdot F + \big[A\cdot\partial b -\partial\bar{c}\cdot Dc\big] + \kappa_4 b\cdot b + \kappa_2\big[2b + g(\bar{c}\times c)\big]\cdot(\bar{c}\times c) +\big[M_A^2 A\cdot A + M_c^2\bar{c}\cdot c\big] \,,
\end{eqnarray}
where
\begin{equation}
     \kappa_2=g/4\alpha\,,\ \ \kappa_4=1/2\alpha\,, \ \ \mathrm{and}\ \ M_A^2=\frac{1}{2}m^2\,, \ \ M_c^2=\frac{1}{\alpha}m^2 \,.
\end{equation}
As it is well known, this Curci--Ferrari Lagrangian is not invariant under the BRST transformation \eqref{BRST} (or \eqref{BRSTx} in more standard form), rather it is invariant under BRST transformation modified by
\begin{equation}
    \delta_m b_a = -\theta m^2 c_a \,,
\end{equation}
which is sometimes denoted as BRSTm invariance. The BRSTm invariance is however not nilpotent causing troubles related to ill defined cohomology. 

There is a suggestive question, whether it is possible to construct the Curci--Ferrari model as a result of the spontaneous BRST symmetry breaking.
We can do it simply by fixing correspondingly the parameters of the effective Lagrangian and then let the BRST be spontaneously broken. It means that in \eqref{Leff} we set, for instance,
\begin{eqnarray}
     (\kappa_3-\kappa_5)+2k_3\langle\varphi\rangle=1\,,\ \ \kappa_1=0\,,\ \ \kappa_2+k_2\frac{g}{2}\langle\varphi\rangle=\frac{g}{4\alpha}\,,\ \ \kappa_4+k_2\langle\varphi\rangle=\frac{1}{2\alpha}\,, \ \ \mathrm{and} \ \ k_2\langle\bar\phi\rangle=\frac{m^2}{\alpha}\,, \ \ k_3\langle\bar\phi\rangle=\frac{m^2}{2} \,.
\end{eqnarray}
The shift of the Fujikawa fields $\varphi\rightarrow\varphi+\langle\varphi\rangle$ and $\bar\phi\rightarrow\bar\phi+\langle\bar\phi\rangle$ allows to rewrite the Lagrangian of the elementary fields, $A$, $c$, $\bar{c}$, and $b$, as the Curci--Ferrari model \eqref{LCF}. Indeed for
\begin{eqnarray}
     &&\langle\bar\phi\rangle=m^2\,, \ \ \langle\varphi\rangle=\frac{1}{2}\,,\\
     &&k_2=\frac{1}{\alpha}\,, \ \ k_3=\frac{1}{2}\,,\ \ \kappa_3-\kappa_5=\frac{1}{2}\,,\ \ \kappa_{1,2,4}=0\,,
\end{eqnarray}
we obtain an equivalent realization of the Curci--Ferrari Lagrangian
\begin{eqnarray}
    \mathcal{L}_\mathrm{eff} &=& \mathcal{L}_\mathrm{F} -\tfrac{1}{4}F\cdot F + \tfrac{1}{2}\big[A\cdot\partial b -\partial\bar{c}\cdot Dc\big] \\
    &&+\big[\tfrac{1}{2} A\cdot A + \tfrac{1}{\alpha}\bar{c}\cdot c\big]\bar\phi + \big[(A\cdot\partial b -\partial\bar{c}\cdot Dc) + \tfrac{1}{\alpha} b\cdot b  +\tfrac{g}{2\alpha}\big(2b + g\bar{c}\times c\big)\cdot(\bar{c}\times c)\big]\varphi  \nonumber\\
    &&+ \big[A\cdot\partial c + \tfrac{1}{\alpha}\big(c\cdot b+\tfrac{g}{2}c\cdot(\bar{c}\times c)\big)\big]\bar\pi - \big[A\cdot\partial \bar{c} + \tfrac{1}{\alpha}\big(\bar{c}\cdot b+\tfrac{g}{2}\bar{c}\cdot \bar{c}\times c\big)\big]\pi \,. \nonumber
\end{eqnarray}

Upon the shift $\varphi\rightarrow\varphi+\langle\varphi\rangle$ and $\bar\phi\rightarrow\bar\phi+\langle\bar\phi\rangle$ we obtain 
\begin{eqnarray}
    \mathcal{L}_\mathrm{eff} &=& \mathcal{L}_\mathrm{CF}+\mathcal{O}(\varphi,\bar\phi,\pi,\bar\pi) \,.
\end{eqnarray}
The Curci--Ferrari model is reproduced by the purely elementary field part of the Lagrangian. The remaining terms containing the Nambu--Goldstone fields are there to guarantee the presence of the BRST symmetry, which is hidden in the sense of the spontaneous symmetry breaking.

\subsection{Derivation of the Curci--Ferrari model together with its modified-BRST symmetry}

In the previous subsection we have derived the Curci--Ferrari Lagrangian as the elementary-field part of the special case of our effective Lagrangian after spontaneous BRST symmetry breaking. The modified-BRST invariance of $\mathcal{L}_\mathrm{CF}$ however emerges just ``accidentally". In our example the modified-BRST invariance is achieved by non-zero values of $(\kappa_3-\kappa_5)$ tuned together with values of $k_2$, $k_3$, and $\langle\varphi\rangle$.

In this subsection we attempt to obtain the Curci--Ferrari Lagrangian and its modified-BRST symmetry both as the consequences of the spontaneous symmetry breaking and at the same time reduce the model parameters to minimum. 
At that aim we extend the BRST transformation, while keeping it still nilpotent, in such manner that after the symmetry breaking the modified-BRST transformation of $b_a$ field arises:
\begin{equation}
    \delta_\mathrm{E}b_a\propto \theta\bar\phi c_a + \dots \ \ \ \stackrel{\mathrm{SSB}}{\longrightarrow}\ \ \ \tilde\delta_\mathrm{E}b_a\propto \theta\langle\bar\phi\rangle c_a + \dots \,,
\end{equation}
where the tilde in $\tilde\delta$ refers to the transformation after the spontaneous symmetry breaking. For such extension the presence of Fujikawa fields is essential. 

The nilpotent extension of the BRST transformation can be done in many ways, but we choose a minimal extension that is serving our purpose and that also fulfills the requirement of the anti-BRST invariance:
\begin{subequations}\label{EBRST}
\begin{eqnarray}
\delta_\mathrm{E}\bar{c}_a &=& \theta \Big[b_a + \bar\pi c_a-\pi \bar{c}_a\Big] \,,\\
\delta_\mathrm{E}b_a &=& \theta\Big[-\bar\phi c_a-\tfrac{g}{2}\bar\pi(c\times c)_a+\bar\pi\pi c_a-\pi b_a\Big] \,.
\end{eqnarray}
\end{subequations}
The rest of the transformations remains standard. The nilpotency can be proved by calculating that
\begin{subequations}\label{EBRSTnilpotency}
\begin{eqnarray}
\delta_\mathrm{E}\Big[b_a + \bar\pi c_a-\pi \bar{c}_a\Big] &=& 0 \,,\\
\delta_\mathrm{E}\Big[-\bar\phi c_a-\tfrac{g}{2}\bar\pi(c\times c)_a+\bar\pi\pi c_a-\pi b_a\Big] &=& 0 \,.
\end{eqnarray}
\end{subequations}
This extension of the BRST transformation amounts to redefine the $b_a$ field
\begin{equation}\label{b_shift}
    b_a\longrightarrow b_a + \bar\pi c_a-\pi \bar{c}_a \,.
\end{equation}
Hence, the extended-BRST invariant effective Lagrangian can be achieved from the previously constructed BRST invariant effective Lagrangian by simple substitution of the $b_a$ field. That however does not help with the necessity to incorporate the $\kappa$ terms. In order to avoid them, we will use the general effective Lagrangian \eqref{Leff_gen} instead of \eqref{Leff}:
\begin{eqnarray}
    \mathcal{L}^\mathrm{gen}_\mathrm{eff}( b\rightarrow b + \bar\pi c-\pi \bar{c}) &=&  -\tfrac{1}{4}F\cdot F +\mathcal{L}_\mathrm{F}  \label{Leff_gen_shift}\\
    && +\big[K_3'(\varphi,\bar\phi)\varphi_3 + K_2'(\varphi,\bar\phi)\varphi_2\big]\bar\phi
    -\big[K_3''(\varphi,\bar\phi)\varphi_3 + K_2''(\varphi,\bar\phi)\varphi_2\big]\bar\pi\pi\nonumber\\
    &&+ \big[2 K_3(\varphi,\bar\phi)\bar\phi_3 + K_2(\varphi,\bar\phi)( \bar\phi_4 +\tfrac{g}{2}\bar\phi_2)\big]  \nonumber\\
    &&+ \big[2K_3'(\varphi,\bar\phi)\pi_3 + K_2'(\varphi,\bar\phi)\big(\pi_4 +\tfrac{g}{2}\pi_2\big)\big]\bar\pi - \big[2K_3'(\varphi,\bar\phi)\bar\pi_3 + K_2'(\varphi,\bar\phi)\big(\bar\pi_4 +\tfrac{g}{2}\bar\pi_2\big)\big]\pi \nonumber\\
    &&-2K_3(\varphi,\bar\phi)\big[\pi_3\bar\pi-\bar\pi_3\pi+(A\cdot c)\partial\bar\pi-(A\cdot\bar c)\partial\pi\big] \nonumber\\
    &&    -2K_2(\varphi,\bar\phi)\big[(\pi_4\bar\pi-\bar\pi_4\pi)+\tfrac{g}{2}(\pi_4\bar\pi-\bar\pi_4\pi)+\varphi_2\bar\pi\pi\big]+2K_2'(\varphi,\bar\phi)\varphi_2\bar\pi\pi\,.\nonumber
\end{eqnarray}
The last two rows emerges from the shift of the $b_a$ field \eqref{b_shift}.

Now, in order to reproduce the Curci--Ferrari model after the field shift \eqref{SSBshift} due to the spontaneous symmetry breaking, we require
\begin{equation}
    \langle\bar\phi\rangle = m^2 \,,
\end{equation}
and
\begin{eqnarray}
    K_3'(\langle\varphi\rangle,\langle\bar\phi\rangle) = \tfrac{1}{2} \,, &&  K_3(\langle\varphi\rangle,\langle\bar\phi\rangle) = \tfrac{1}{2}  \,,\\
    K_2'(\langle\varphi\rangle,\langle\bar\phi\rangle) = \tfrac{1}{\alpha} \,, &&  K_2(\langle\varphi\rangle,\langle\bar\phi\rangle) = \tfrac{1}{2\alpha}  \,.
\end{eqnarray}
If we require that these relations are valid for all values of $\langle\varphi\rangle$ and if we avoid the dependence of $K_{2,3}$ on $\bar\phi$ for simplicity, then we can fix the functional form of $K_{2,3}$ to
\begin{eqnarray}
    K_3(\varphi,\bar\phi) &=& \tfrac{1}{2}\mathrm{e}^{\varphi-\langle\varphi\rangle} \,=\, K_3' \,=\, K_3''\,,\\
    K_2(\varphi,\bar\phi) &=& \tfrac{1}{2\alpha}\mathrm{e}^{2(\varphi-\langle\varphi\rangle)} \,=\, \tfrac{1}{2}K_2' \,=\, \tfrac{1}{4}K_3''\,.
\end{eqnarray}
Feeding this into \eqref{Leff_gen_shift} we get
\begin{eqnarray}
    \mathcal{L}^\mathrm{gen}_\mathrm{eff}( b\rightarrow b + \bar\pi c-\pi \bar{c}) &=&  -\tfrac{1}{4}F\cdot F +\mathcal{L}_\mathrm{F}  \label{Leff_gen_shift_exp}\\
    && +\big[K_3(\varphi)\varphi_3 + 2K_2(\varphi)\varphi_2\big](\bar\phi-\bar\pi\pi)
       +\big[2 K_3(\varphi)\bar\phi_3 + K_2(\varphi)( \bar\phi_4 +\tfrac{g}{2}\bar\phi_2)\big] \nonumber\\
    &&-2K_3(\varphi)\big[(A\cdot c)\partial\bar\pi-(A\cdot\bar c)\partial\pi\big] \nonumber\,,
\end{eqnarray}
with seemingly miraculous cancellation of the $(\pi_i\bar\pi-\bar\pi_i\pi)$ terms. After the spontaneous BRST symmetry breaking and the field shift \eqref{SSBshift} we reproduce the Curci--Ferrari model
\begin{eqnarray}
    \mathcal{L}^\mathrm{gen}_\mathrm{eff}( b\rightarrow b + \bar\pi c-\pi \bar{c}) &=&  \mathcal{L}_\mathrm{CF} +\mathcal{L}_\mathrm{F}  \label{Leff_gen_shift_exp_CF}\\
    && +\big[\tfrac{m^2}{2}\big(\mathrm{e}^{\varphi}-1\big)\varphi_3 + \tfrac{m^2}{\alpha}\big(\mathrm{e}^{2\varphi}-1\big)\varphi_2\big]
    +\big[\tfrac{1}{2}\mathrm{e}^{\varphi}\varphi_3 + \tfrac{1}{\alpha}\mathrm{e}^{2\varphi}\varphi_2\big](\bar\phi-\bar\pi\pi)\nonumber\\
    &&+\big[\big(\mathrm{e}^{\varphi}-1\big)\bar\phi_3 + \tfrac{1}{2\alpha}\big(\mathrm{e}^{2\varphi}-1\big)( \bar\phi_4 +\tfrac{g}{2}\bar\phi_2)\big] -\mathrm{e}^{\varphi}\big[(A\cdot c)\partial\bar\pi-(A\cdot\bar c)\partial\pi\big] \nonumber\,,
\end{eqnarray}
together with the modified-BRST invariance
\begin{subequations}\label{EBRST_SSB}
\begin{eqnarray}
\tilde\delta_\mathrm{E}\bar{c}_a &=& \theta \Big[b_a + \bar\pi c_a-\pi \bar{c}_a\Big] \,,\\
\tilde\delta_\mathrm{E}b_a &=& \theta\Big[-m^2 c_a-\bar\phi c_a-\tfrac{g}{2}\bar\pi(c\times c)_a+\bar\pi\pi c_a-\pi b_a\Big] \,.
\end{eqnarray}
\end{subequations}


\section{Conclusions}

In this paper, we present a new perspective on spontaneous BRST symmetry breaking that relies on an analogy with the Chiral Quark Model \cite{Manohar1984}, which describes spontaneous chiral symmetry breaking through the quark condensate. Our approach elaborates on Fujikawa’s framework \cite{Fujikawa1983}, in which spontaneous BRST symmetry breaking is introduced via a condensing scalar BRST--anti-BRST quartet. We propose that spontaneous BRST symmetry breaking in the Yang--Mills quantum field theory is provided by the condensation of the Fujikawa fields that are interpreted as composites of the elementary gluon and ghost fields. This transition effectively describes the shift from the deconfined ultraviolet regime to the infrared, where strong interactions lead to the formation of colorless hadrons and the suppression of color-carrying partons as asymptotic states.

We propose that spontaneous BRST symmetry breaking is accompanied by simultaneous spontaneous anti-BRST symmetry breaking, thereby justifying the presence of two Fujikawa Nambu--Goldstone modes. In direct analogy with how pions in the non-linear representation couple to quarks in the Chiral Quark Model, we couple the Fujikawa quartet to the field content of the Yang--Mills quantum field theory. The central outcome is a BRST- and anti-BRST-invariant effective Lagrangian formulated generally to all orders in the Fujikawa fields. Within this framework, the dynamical generation of effective mass terms for gluons and ghosts emerges as a fundamental nonperturbative consequence of the non-zero vacuum expectation values of the Fujikawa fields. These masses provide a physical screening mechanism for long-wavelength fluctuations, which is consistent with lattice observations showing the infrared saturation of the gluon propagator to a finite, non-zero value \cite{Binosi2026}. 

Because models of massive gluons has proven to be an effective way of reproducing lattice results, recent work \cite{vanEgmond2026} proposed the idea to study the Curci-Ferrari model in the Maximal-Abelian gauge (MAG). Our proposal is somewhat orthogonal to that. We demonstrate the possibility of reproducing the Curci--Ferrari model of massive Yang--Mills fields from spontaneously broken effective Lagrangian. To reproduce not only the Lagrangian but also the Curci--Ferrari modified-BRST invariance, we introduce an extended-BRST transformation which, upon spontaneous symmetry breaking, reduces at linear order in fields to the modified-BRST transformation. The non-linear terms of the transformation — whose very existence fundamentally relies on the Fujikawa fields — ensure the nilpotence of the underlying invariance, now hidden in the sense of spontaneous symmetry breaking. By providing a nilpotent foundation for the Curci--Ferrari model, our model resolves the historical lack of nilpotency that has long been considered a source of non-unitarity and gauge dependence in massive theories \cite{Lavrov2012}. Furthermore, our approach ensures that the theory remains consistent with the background field method, allowing for the definition of a unique, process-independent effective charge that remains well-defined down to the zero-momentum limit \cite{Binosi2003,Aguilar2009}. It remains to study the Curci-Ferrari model by the inclusion of matter fields.

By this initiative, we aim to open a new line of research in which much still needs to be done to test its viability and relevance. Technically, our extension of Yang--Mills quantum field theory by the Fujikawa quartet is similar to the Gribov--Zwanziger extension by a quartet of auxiliary vector and colored fields. The purpose of the Gribov--Zwanziger extension is to localize the physical field sector to the first Gribov region upon integrating out the auxiliary fields. As a result, in the Gribov--Zwanziger model the BRST invariance is broken, but not spontaneously in the narrow sense of a phase transition as elaborated in this paper. Our proposal opens the possibility that the Gribov problem -- the proliferation of gauge copies -- may be naturally mitigated by the dynamical mass generation itself \cite{Gao2018}. The infrared mass serves to screen the fluctuations that would otherwise drive field configurations toward the Gribov horizon, effectively disfavoring the near-horizon configurations where the Faddeev-–Popov determinant changes sign. It is of course mandatory to address the Gribov problem within the model of spontaneous BRST symmetry breaking proposed here in more detail. If needed, in principle, it is conceivable to further extend the present effective model by additional composite quartets serving as the Gribov--Zwanziger fields, in an analogous way to how vector hadron resonances are added to the Chiral Quark Model; we leave this to future work. 

The ultimate question of color confinement is left for future work, though our framework provides a promising path for its resolution. Confinement can be rigorously understood through the Kugo--Ojima criterion, which requires the Kugo--Ojima function to saturate at the value $-1$ in the infrared limit for the existence of a well-defined color charge that annihilates physical states \cite{Kugo1979,Kugo1995}. While lattice data suggests that the Kugo--Ojima function approaches to but does not reach this limit, our model’s inclusion of massless Nambu–-Goldstone modes might offer a mechanism for the ``quartet mechanism" which removes unphysical longitudinal and ghost degrees of freedom from the physical spectrum. The Fujikawa massless Nambu-Goldstone modes should be introduced manually into the lattice framework, because otherwise their effect can be missed as their existence in a theory relies on space-time being a continuum. It appears appealing to address color confinement in terms of a spontaneous-symmetry-breaking deformation of an appropriately extended BRST transformation \cite{Amaral2022}. In the broken phase, the massive gauge fields appear in non-trivial BRST cocycles through combinations with would-be Goldstone bosons, effectively removing transverse gluons from the physical cohomology and ensuring that all observable states are color singlets.

We hope this work prepares the ground for further contributions to the ongoing effort to understand the nonperturbative dynamics of strong interactions.

\begin{acknowledgments}
We acknowledge the support of A.S. by the Ministry of Education, Youth and Sport of the Czech Republic within International mobilities of researchers on CTU — VTA under the contract number
CZ.02.2.69/0.0/0.0/18\_053/0016980 that allowed us to accidentally meet in Bogot\'a and initiate this exciting project. This research activity is part of the project ``Aspects of spontaneous BRST symmetry breaking in Yang-Mills theories", code 60980, for a joint collaboration of the research group ``Grupo de Campos y Particulas" of Universidad Nacional de Colombia in Bogot\'a, MINCIENCIAS COL0007847, and the Institute of Experimental and Applied Physics, Czech Technical University in Prague. A.R.F. is extremely grateful for the kind hospitality and financial support received by the Institute of Experimental and Applied Physics, Czech Technical University in Prague.
 
\end{acknowledgments}

\appendix

\section{BRST and anti-BRST transformations of the fields}\label{A1}

We denote the BRST transformation of the field $\psi$ as $\delta\psi$, and the anti-BRST transformation as $\bar\delta\psi$. For the elementary fields $A^\mu_a$, $c_a$, $\bar{c}_a$ and $b_a$, and for the Fujikawa fields $\varphi$, $\pi$, $\bar\pi$ and $\bar{\phi}$ we have the following BRST and anti-BRST transformations:
\begin{eqnarray}\label{BRSTx}
    \begin{array}{rclrclrclrcl}
        \vphantom{\Big(}\delta A^\mu_a &=& \theta(D^\mu c)_a\,, & 
        \bar\delta A^\mu_a &=& \bar\theta(D^\mu \bar{c})_a\,, &
        \delta\varphi &=&\theta\pi \,, \quad\quad&
        \bar\delta\varphi &=&\bar\theta\bar\pi \,,\\
        \vphantom{\Big(}\delta c_a &=& -\frac{g}{2}\theta(c\times c)_a \,, \quad\quad& 
        \bar\delta \bar{c}_a &=& -\frac{g}{2}\bar\theta(\bar{c}\times \bar{c})_a \,,&
        \delta\pi&=&0 \,,&
        \bar\delta\bar\pi&=&0 \,,\\
        \vphantom{\Big(}\delta \bar{c}_a &=& \theta b_a \,, &
        \bar\delta c_a &=& \bar\theta\, \bar{b}_a = \bar\theta\big[-b_a- g(\bar{c}\times c)_a\big] \,,\quad\quad&
        \delta\bar\pi &=&\theta\bar\phi \,,\quad&
        \bar\delta\pi &=&\bar\theta\phi\,=\,-\bar\theta\bar\phi \,,\\
        \vphantom{\Big(}\delta b_a &=& 0 \,, & \bar\delta \bar{b}_a &=& 0 \,, \quad& \delta\bar\phi&=& 0\,, & \bar\delta\phi&=& 0\,.
    \end{array}
\end{eqnarray}
The (anti-)commutator of the elementary fields with the BRST charge operator $Q$ is
\begin{subequations}\label{BRST}
\begin{eqnarray}
\{Q,c_a\} &=& -\frac{g}{2}(c\times c)_a \,,\\
\{Q,\bar{c}_a\} &=& b_a \,,\\
{}[Q,A^\mu_a] &=& D^\mu c_a\,, \\
{}[Q,b_a] &=& 0 \,.
\end{eqnarray}    
\end{subequations}
The BRST charge $Q$ as well as the ghost, $c_a$, and anti-ghost, $\bar{c}_a$, fields satisfy the Fermi statistics.

Apart from the proper BRST symmetry there exists also symmetry of ghost number conservation (GNC) with the generator $D$ satisfying the (anti-)commutation relations \cite{Fujikawa1983,Bonora1980,Bonora2021}
\begin{eqnarray}
{}[D,c_a] &=& \I c_a \,,\\
{}[D,\bar{c}_a] &=& -\I \bar{c}_a \,,\\
{}[D,A^\mu_a] &=& 0 \,, \\
{}[D,b_a] &=& 0 \,, \\
{}[Q,D] &=& \I Q  \,.
\end{eqnarray}

\section{BRST and anti-BRST transformations of the composite operators}

We list here all of the composite operators that we are using in this work:
\begin{eqnarray}
    \begin{array}{rclrclrcl}
    \vphantom{\Big(}\varphi_2&\equiv& \bar{c}\cdot c & &&&&& \\
    \vphantom{\Big(}\varphi_3&\equiv& A\cdot A & &&&&& \\
    &&&&&&&& \\
    \vphantom{\Big(}\pi_1 &\equiv& c\cdot(A\otimes A) \,,\quad\quad & 
    \eta_1 &\equiv& 2 A\cdot(c\otimes Dc) - \frac{g}{2}(A\otimes A)\cdot(c\times c) \,, \quad\quad&
    \bar{\phi}_1 &\equiv& b\cdot(A\otimes A) - 2A\cdot(\bar{c}\otimes Dc)  \,,\\
    \vphantom{\Big(}\pi_2 &\equiv& c\cdot(\bar{c}\times c) \,, &
    \eta_2 &\equiv& b\cdot(c\times c) \,, &
    \bar{\phi}_2 &\equiv& 2b\cdot(\bar{c}\times c) + g(\bar{c}\times c)\cdot(\bar{c}\times c) \,,\\
    \vphantom{\Big(}\pi_3 &\equiv& A\cdot\partial c \,, &
    \eta_3 &\equiv& 0  \,, &
    \bar{\phi}_3 &\equiv& A\cdot\partial b -\partial\bar{c}\cdot Dc \,,\\
    \vphantom{\Big(}\pi_4 &\equiv& c\cdot b \,, &
    \eta_4 &\equiv& -\frac{g}{2}b\cdot(c\times c) \,, &
    \bar{\phi}_4 &\equiv& b\cdot b \,,\\
    \vphantom{\Big(}\pi_5 &\equiv& c\cdot \partial A \,, &
    \eta_5 &\equiv& \partial\big(c\cdot\partial c + \frac{g}{2}A\cdot(c\times c)\big) \,, &
    \bar{\phi}_5 &\equiv& b\cdot\partial A -\bar{c}\cdot\partial Dc \,.\\
    \end{array}
\end{eqnarray}
Their BRST and anti-BRST transformations are:
\begin{eqnarray}
    \begin{array}{rclrcl}
        \vphantom{\Big(}\delta\varphi_2 &=&\theta\big[\pi_4+\frac{g}{2}\pi_2\big] \,, \quad\quad& 
        \bar\delta\varphi_2 &=&\bar\theta\big[\bar\pi_4+\frac{g}{2}\bar\pi_2\big] \,,\\
        \vphantom{\Big(}\delta\varphi_3 &=&2\theta\pi_3 \,,\quad\quad& 
        \bar\delta\varphi_3 &=&2\bar\theta\bar\pi_3 \,,\\
        \vphantom{\Big(}\delta\pi_i &=&\theta \eta_i \,, &
        \bar\delta\bar\pi_i &=&-\bar\theta\bar\eta_i \,,\\
        \vphantom{\Big(}\delta\bar\pi_i &=&\theta\bar\phi_i \,, & \bar\delta\pi_i &=&-\bar\theta\bar\phi_i\,,\\
        \vphantom{\Big(}\delta\bar\phi_i &=&0 \,, & \bar\delta\bar\phi_i &=&0\,
    \end{array}
\end{eqnarray}


where
\begin{equation}
\bar\eta_i\equiv\eta_i\Big|_{c\rightarrow\bar c}\quad\mathrm{and}\quad\bar\pi_i\equiv\pi_i\Big|_{c\rightarrow\bar c}\,.
\end{equation}

As a matter of the nilpotence of both $\delta$ and $\bar\delta$ applied to $\varphi_2$ and $\varphi_3$ there are relations
\begin{eqnarray}
    \eta_3=0\,,\quad \eta_4=-\frac{g}{2}\eta_2 \,,
\end{eqnarray}
and so for $\bar\eta$'s. No such relation was found for $\eta_1$ nor for $\eta_5$, as a consequence of the fact that neither $\pi_1$ nor $\pi_5$ are transforms of any other field.

\section{BRST Superfield formalism}\label{App_Superfields}

The superfield formalism is based on the extension of the set of space-time coordinates by a real Grassmannian coordinate $\theta$ \cite{Fujikawa1983,Bonora1980,Bonora2021} which carries the ghost number $d=-1$ and $\mathrm{dim}=-1$. The Taylor series in $\theta$ of the superfield $\psi(x,\theta)$ has, due to the Grassmannian nature of $\theta$, just two terms,  
\begin{equation}
    \psi(x,\theta)=\psi(x) + \theta \delta\psi(x) \,,
\end{equation}
hence referred to as the superfield {\it doublet}, where
\begin{equation}
    \delta\psi(x)=\frac{\partial_L}{\partial\theta}\psi(x,\theta)\Big|_{\theta=0} \,.
\end{equation}

The superfield doublet $\psi(x,\theta)$ has then general transformation properties\footnote{notice the minus sign in the transformation of $\theta$ opposite to \cite{Fujikawa1983} where we consider it as a typo}:
\begin{eqnarray}
    \mathrm{e}^{\lambda Q}\psi(x,\theta)\mathrm{e}^{-\lambda Q}=\psi(x,\theta+\lambda) \,,\\
    \mathrm{e}^{\I\rho D}\psi(x,\theta)\mathrm{e}^{-\I\rho D}=\mathrm{e}^{d\rho}\psi(x,\mathrm{e}^{-\rho}\theta) \,,
\end{eqnarray}
where $d$ is the ghost number of the superfield and $\lambda$ and $\rho$ are transformation parameters of the proper BRST $Q$ and ghost number $D$, respectively.

The elementary fields $A^a_\mu$, $c^a$, $\bar{c}^a$, and $b^a$ are components of their own  of ghost number $d$ and mass dimension $\dim$:
\begin{equation}
\begin{array}{c|c|l}
\mathrm{ghost\ number} & \mathrm{mass\ dimension} & \mathrm{superfield} \\ \hline\hline
\vphantom{\Big(} 0 & 1 &  {\cal{A}}^a_\mu(x,\theta) = A^a_\mu(x)+\theta\big(\partial_\mu c^a(x)- g f^{abc}A_\mu^b(x) c^c(x)\big) \\ 
\vphantom{\Big(}+1 & 1 &  {\cal{C}}^a(x,\theta) = c^a(x)-\theta\frac{g}{2}f^{abc}c^b(x) c^c(x) \\ 
\vphantom{\Big(}-1 & 1 &  \bar{\cal{C}}^a(x,\theta) = \bar{c}^a(x)+\theta b^a(x) \\ 
\vphantom{\Big(}0 & 2 &  {\cal{B}}^a(x,\theta)= \partial_\theta\bar{\cal{C}}^a(x,\theta) = b^a(x) \\ \hline
\end{array}
\end{equation}

The color-singlet superfield containing the information of \eqref{Qbarpi} is 
\begin{equation}
\begin{array}{c|c|l}
\mathrm{ghost\ number} & \mathrm{mass\ dimension} & \mathrm{superfield} \\ \hline\hline 
\vphantom{\Big(}-1 & 3 &  \xi_1(x,\theta)\equiv d_{abc}\bar{\cal{C}}_a(x,\theta){\cal{A}}^\mu_b(x,\theta) {\cal{A}}_{\mu c}(x,\theta) = \bar{\pi}_1 + \theta \bar{\phi}_1 \\ 
\vphantom{\Big(}-1 & 3 &  \xi_2(x,\theta)\equiv f_{abc}\bar{\cal{C}}_a(x,\theta)\bar{\cal{C}}_b(x,\theta){\cal{C}}_c(x,\theta) = \bar{\pi}_2 + \theta\bar{\phi}_2 \\ 
\vphantom{\Big(}-1 & 3 &  \xi_3(x,\theta)\equiv {\cal{A}}^\mu_a(x,\theta)\partial_\mu\bar{\cal{C}}_a(x,\theta) = \bar{\pi}_3 + \theta\bar{\phi}_3 \\ 
\vphantom{\Big(}-1 & 3 &  \xi_4(x,\theta)\equiv \bar{\cal{C}}_a(x,\theta){\cal{B}}_a(x,\theta) = \bar{\pi}_4 + \theta\bar{\phi}_4 \\ 
\vphantom{\Big(}-1 & 3 &  \xi_5(x,\theta)\equiv \bar{\cal{C}}_a(x,\theta)\partial_\mu{\cal{A}}^\mu_a(x,\theta) = \bar{\pi}_5 + \theta\bar{\phi}_5 \\ \hline
\end{array}
\label{xi}
\end{equation}
where $v$ is a mass parameter by which we can normalize the composite superfield. Additionally, this field can be complemented by two other $d=0$ color-singlet superfield doublets

\begin{equation}
\begin{array}{c|c|l}
\mathrm{ghost\ number} & \mathrm{mass\ dimension} & \mathrm{superfield} \\ \hline\hline
\vphantom{\Big(}0 & 2 &  \varphi_2(x,\theta)\equiv \bar{\cal{C}}_a{\cal{C}}_a = \varphi_2 + \theta\big(\pi_4+\frac{g}{2}\pi_2\big) \\ 
\vphantom{\Big(}0 & 2 &  \varphi_3(x,\theta)\equiv {\cal{A}}^\mu_a{\cal{A}}^{\mu a} = \varphi_3 + 2\theta\pi_3 \\ \hline
\end{array}
\label{varphi}
\end{equation}

The Fujikawa fields can be embedded into the color-singlet superfields $\xi(x,\theta)$ and $\varphi(x,\theta)$
\begin{equation}
\begin{array}{c|c|l}
\mathrm{ghost\ number} & \mathrm{mass\ dimension} & \mathrm{superfield} \\ \hline\hline 
\vphantom{\Big(}-1 & 1 &  \xi(x,\theta)\equiv \bar{\pi}(x) + \theta \bar{\phi}(x) \\ 
\vphantom{\Big(}0 & 0 &  \varphi(x,\theta)\equiv \varphi(x) + \theta \pi(x) \\ \hline
\end{array}\label{Ff}
\end{equation}
Their interpolating superfield operators are some linear combination of $\xi_i(x,\theta)$ and $\varphi_j(x,\theta)$ from \eqref{xi} and \eqref{varphi}.

\bibliography{biblio}

\end{document}